# The rotating Morse potential model for diatomic molecules in the tridiagonal J-matrix representation: I. Bound states

I. Nasser, M. S. Abdelmonem, and H. Bahlouli
*Physics Department, King Fahd University of Petroleum & Minerals, Dhahran 31261, Saudi Arabia*

A. D. Alhaidari
*Shura Council, Riyadh 11212, Saudi Arabia*

This is the first in a series of articles in which we study the rotating Morse potential model for diatomic molecules in the tridiagonal J-matrix representation. Here, we compute the bound states energy spectrum by diagonalizing the finite dimensional Hamiltonian matrix of $H_2$, LiH, HCl and CO molecules for arbitrary angular momentum. The calculation was performed using the J-matrix basis that supports a tridiagonal matrix representation for the reference Hamiltonian. Our results for these diatomic molecules have been compared with available numerical data satisfactorily. The proposed method is handy, very efficient, and it enhances accuracy by combining analytic power with a convergent and stable numerical technique.

## I. INTRODUCTION

The Morse potential plays a dominant role in describing the interaction of atoms in diatomic and even polyatomic molecules [1-3]. An effective potential, which is the sum of the centrifugal potential term that depend on the angular momentum $\ell$ and the Morse potential, has been used as a model for such interactions. It is referred to as the rotating Morse potential. For $\ell = 0$, the radial Schrödinger equation with this potential has been solved exactly [1]. However, for $\ell > 0$ only numerical solutions are possible where several approximation techniques have been proposed and extensively used with varying degrees of accuracy and stability [4-19].

Numerical solutions, with and without Pekeris approximation, have been used to calculate the energy spectrum [4-7]. The Pekeris approximation [4] is based on the expansion of the centrifugal part in exponential terms with exponents that depend on an inter-nuclear distance parameter. This is why Pekeris approximation is valid only for very small spatial variations from the inter-nuclear separation (i.e. for low vibrational energy). Other methods that have also been used include the variational method with Pekeris approximation [8], supersymmetry (SUSY) [9,10], the hyper-virial (HV) perturbation method with the full potential without Pekeris approximation [11], the shifted $1/N$ expansion (SE) [12] and the modified shifted large $1/N$ approach (MSE) [13]. In [13] it was stated that the MSE method leads to the exact wavefunction for pure vibrational states and, as such, gives the most precise results. However, it is cumbersome and requires a lot of computational time.

Other methods, that are semi-analytic, have also been used. These include the Nikiforov–Uvarov (NU) method [14,15] and the asymptotic iteration methods (AIM) [16]. In the AIM method, the energy eigenvalues are obtained by simple transformation of the radial Schrödinger equation and the wavefunction is calculated iteratively. The two-point quasi-rational approximation technique (TQA) [17], which is considered as an extension of the Padé procedure, was used for the hydrogen molecule. The exact quantization rule (EQR) [18] and the Fourier grid Hamiltonian method (FGH) [19] have also been used in this context.

Our approach for the study of the rotating Morse potential model was inspired by the J-matrix method [20]. It is an algebraic method for extracting bound states and scattering information using computational tools devised entirely in square integrable bases. Physically bound states are expandable in square integrable ($L^2$) functions that are compatible with the domain of the Hamiltonian. In fact, $L^2$ functions arise naturally as negative energy eigenfunctions for the discrete bound states such as in the familiar hydrogen atom. Hence it seems natural that the use of discrete basis sets offers considerable advantage in the calculation of bound states since this scheme requires only standard matrix technique rather than the usual approach of numerical integration of the differential equations. In this approach, when searching for algebraic or numerical bound states solutions, the wave function $\psi$ spans the space of square integrable functions with discrete basis elements $\{\phi_n\}_{n=0}^{\infty}$. That is $|\psi(\vec{r},E)\rangle = \sum_n f_n(E)|\phi_n(\vec{r})\rangle$, where $\vec{r}$ is the set of coordinates for real space, and $E$ is the system's energy. The basis functions must be compatible with the domain of the Hamiltonian and satisfy the boundary conditions. In our case, this means that the basis set must vanish at $r = 0$ and $r = \infty$. Typically, when calculating the discrete spectrum, the choice of basis is limited to those that carry a diagonal representation of the Hamiltonian $H$. That is, one looks for an $L^2$ basis set $\{\phi_n\}_{n=0}^{\infty}$ such that $H|\phi_n\rangle = E_n|\phi_n\rangle$ giving the discrete spectrum of $H$. The continuous spectrum is obtained from the analysis of an infinite sum of these *complete* basis functions.

Solving the eigenvalue problem requires an efficient method for computing matrix elements of the Hamiltonian. In this article, we relax the restriction of a diagonal matrix representation for the Hamiltonian. Instead, we split our original Hamiltonian in two parts as $H = H_0 + V$ where $H_0$ is the part of the Hamiltonian that could be treated analytically while the remaining part, $V$, is to be handled numerically. In dealing with the matrix representation of $H_0$, we do not require it to be diagonal but only that it is hermitian and tridiagonal. That is, the matrix elements of the reference wave operator should take the following form

$$\langle \phi_n | H_0 - E | \phi_m \rangle = (a_n - z)\delta_{n,m} + b_n \delta_{n,m-1} + b_{n-1}\delta_{n,m+1}, \qquad (1.1)$$

where $z$ and the coefficients $\{a_n, b_n\}_{n=0}^{\infty}$ are real and, in general, functions of the energy, angular momentum, and potential parameters. Thus, this approach embodies the analytical power in dealing with the reference $H_0$ problem as well as the very accurate numerical quadrature approach [21] used in the evaluation of the potential matrix elements. In addition, we benefit from the existence of two different bases that support a tridiagonal matrix representation of the reference Hamiltonian. Each one has two parameters which will be tuned to enhance convergence, stability, and of course accuracy. For a given potential model, the choice of basis is made with an eye on these three computational characteristics.

The paper is organized as follows: In section II, we review and summarize the theoretical formalism of our approach. Then, in section III, we develop a numerical procedure for implementing our approach on the S-wave Morse potential. This enables us to measure the accuracy of our numerical results against analytic ones for four different types of binary molecules. In section IV, we apply our approach to the calculation of the energy eigenvalues for the rotating Morse potential where $\ell \neq 0$ and compare our results with those obtained by other methods elsewhere. This will be followed, in section V, by a short summary of our findings and a conclusion of the work. In future articles, we will develop an efficient method for locating resonances in these models using the same representation together with the complex scaling method. We will also investigate generalized three-parameter Morse potential models that are more suitable and give better description of various molecules. Moreover, the scattering matrix in the J-matrix method will be employed to enhance the accuracy of computations at low basis dimension.

## II. THEORETICAL FORMALISM

As explained in the introduction the real power of our approach is that: (1) it allows for the analytic computation of the matrix elements of the reference Hamiltonian, $H_0$, and (2) it provides for a highly accurate evaluation of the potential matrix elements using Gauss quadrature approximation. The computation of the eigenvalues of the resulting full Hamiltonian is performed to the desired accuracy using two free parameters, a length scale parameter $\lambda$ and the dimension of the basis space $N$.

In the atomic units, $\hbar = m = 1$, the radial time-independent Schrödinger equation for a system with reduced mass $m$ in a spherically symmetric potential $V(r)$ reads as follows

$$(H-E)|\psi\rangle = \left[-\frac{1}{2}\frac{d^2}{dr^2} + \frac{\ell(\ell+1)}{2r^2} + V(r) - E\right]|\psi\rangle = 0, \quad (2.1)$$

where $|\psi\rangle$ is the wavefunction which is parameterized by the angular momentum $\ell$, energy $E$, and the potential parameters. The continuum could be discretized by taking $|\psi\rangle$ as an element in an $L^2$ space with a complete basis set $\{\phi_n\}$. The integration measure in this space is $dr$. We parameterize the basis by a positive length scale parameter $\lambda$ as $\{\phi_n(\lambda r)\}$. The following choice of basis functions is compatible with the domain of the above Hamiltonian and satisfies the boundary conditions (the vanishing of the wave function at $r = 0$ and $r \to \infty$)

$$\phi_n(x) = A_n x^{\alpha} e^{-x/2} L_n^{\nu}(x); \qquad n = 0, 1, 2, .. \quad (2.2)$$

where $x = \lambda r$, $\alpha > 0$, $\nu > -1$, $L_n^{\nu}(x)$ is the associated Laguerre polynomial of order $n$, and $A_n$ is the normalization constant $\sqrt{\lambda \Gamma(n+1)/\Gamma(n+\nu+1)}$. The matrix representation of the reference Hamiltonian $H_0$ ($\equiv H - V$) in this basis is written as

$$(H_0)_{nm} = \langle \phi_n(x) | -\frac{\lambda^2}{2}\frac{d^2}{dx^2} + \frac{\lambda^2}{2}\frac{\ell(\ell+1)}{x^2} | \phi_m(x) \rangle. \quad (2.3)$$

In the manipulation of (2.3) we use the differential equation, differential formulas, and three-term recursion relation of the Laguerre polynomials [22]. As a result, we obtain the following elements of the matrix representation of the reference Hamiltonian, with $2\alpha = \nu + 1$:

$$\left(H_0\right)_{nm} = \tfrac{\lambda^2}{8}(2n+\nu+1)\delta_{n,m} + \tfrac{\lambda^2}{8}\sqrt{n(n+\nu)}\delta_{n,m+1} + \tfrac{\lambda^2}{8}\sqrt{(n+1)(n+\nu+1)}\delta_{n,m-1} \quad (2.4)$$

This simple tridiagonal matrix representation for $H_0$ holds for $\nu = 2\ell + 1$. The basis $\{\phi_n\}$, on the other hand, is not orthogonal. Its overlap matrix,

$$\langle \phi_n | \phi_m \rangle \equiv \Omega_{nm} = (2n+\nu+1)\delta_{n,m} - \sqrt{n(n+\nu)}\delta_{n,m+1} - \sqrt{(n+1)(n+\nu+1)}\delta_{n,m-1}, \quad (2.5)$$

is tridiagonal. Now, the only remaining quantities that are needed to perform the calculation are the matrix elements of the potential $V(r)$. These are written as

$$\begin{aligned} V_{nm} &= \int_0^\infty \phi_n(\lambda r) V(r) \phi_m(\lambda r) dr \\ &= A_n A_m \int_0^\infty x^\nu e^{-x} L_n^\nu(x) L_m^\nu(x) [xV(x/\lambda)] dx \end{aligned} \quad (2.6)$$

The evaluation of such an integral is almost always done numerically. In our case, we use the Gauss quadrature approximation [21] giving

$$V_{nm} \cong \sum_{k=0}^{N-1} \Lambda_{nk} \Lambda_{mk} [\varepsilon_k V(\varepsilon_k/\lambda)] \quad (2.7)$$

for an adequately large integer $N$, and where $\varepsilon_k$ and $\{\Lambda_{nk}\}_{n=0}^{N-1}$ are the respective $N$ eigenvalues and eigenvectors of the $N \times N$ tridiagonal symmetric matrix $J$, whose elements are

$$J_{n,n} = 2n+\nu+1, \quad J_{n,n+1} = -\sqrt{(n+1)(n+\nu+1)}. \quad (2.8)$$

where $\nu = 2\ell + 1$ for the "Laguerre basis" (2.2). There is another basis (known as the "oscillator basis") that carries a tridiagonal matrix representation for the same reference Hamiltonian. This along with the corresponding matrix manipulations are presented in Appendix. A.

The reference Hamiltonian $H_0$ in this representation could therefore be fully accounted for analytically, whereas the potential $V$ is approximated by its representation in a subset of the basis, such that

$$H_{nm} \cong \begin{cases} (H_0)_{nm} + V_{nm} & ; \quad n,m \leq N-1 \\ 0 & ; \quad n,m > N-1 \end{cases}. \quad (2.9)$$

This representation is the fundamental underlying structure of the $J$-matrix method [20]. It is an algebraic method that has the advantage of taking into account the full contribution of $H_0$ exactly without any truncation. Nevertheless, we will confine our implementation of the present approach to the finite matrix representation (in the subspace $\{\phi_n\}_{n=0}^{N-1}$) for the potential $V$ as well as for the reference Hamiltonian $H_0$. Taking into account the full reference Hamiltonian should result in a substantial improvement on the accuracy of the results. This is currently being pursued in dealing with resonances and will be reported in the near future. Now, in our numerical implementation we use the Morse potential model for diatomic molecules defined by

$$V(r) = D_e \left[ e^{-2\alpha(r-r_e)/r_e} - 2e^{-\alpha(r-r_e)/r_e} \right] \quad (2.10)$$

where $D_e$ is the dissociation energy, $r_e$ is the equilibrium inter-nuclear distance, and $\alpha$ is a parameter that controls the width of the potential well. The numerical values of these parameters are shown in Table 1 for different diatomic molecules along with the sources from which these data were extracted.

In the following sections, we make explicit calculations of the bound states for the Morse potential with finite angular momentum using this $L^2$ basis representation. Our calculation strategy will be as follows: first we choose a basis set and study the stability of the eigenvalues as we vary the scaling parameter $\lambda$ until we reach a plateau in $\lambda$. Then, for a selected value of $\lambda$ from within the plateau, we compute the eigenvalues of the full Hamiltonian and compare them with available numerical results for the given molecule. For $\ell = 0$, the comparison is made against well-known exact analytic results [3,8]. To improve the numerical results further we increase the dimension of the space $N$ until the desired accuracy is reached. The interplay between these two parameters, $\lambda$ and $N$, constitutes the power of our approach and results in fast converging computations with high stability for relatively small values of $N$.

### III. S-WAVE MORSE POTENTIAL

The S-wave ($\ell = 0$) Morse potential has been solved analytically [3,8] and its bound states energy is given in a closed form. A brief derivation of this analytic result will be given below for two reasons. On the one hand, it is a very valuable ingredient since it constitutes a reliable reference and calibration tool to test the accuracy of our numerical approach. On the other hand, it brings out an alternative three-parameter model based on a *generalized* Morse potential that has a definite advantage over the standard two-parameter model. If we consider the case $\ell = 0$ in equation (2.1), then the S-wave Schrödinger equation for the *generalized* Morse potential reads as follows

$$\left\{ -\frac{\hbar^2}{2m}\frac{d^2}{dr^2} + D_e \left[ e^{-2\alpha\left(\frac{r}{r_e}-1\right)} - 2\beta e^{-\alpha\left(\frac{r}{r_e}-1\right)} \right] - E \right\} \chi = 0, \qquad (3.1)$$

where the potential strength $D_e > 0$, the parameters $\alpha$ and $r_e$ are positive and $\beta$ is a deformation parameter of the classical Morse potential. This equation has an exact analytic solution for the bound states energy spectrum. Its derivation goes as follows. If we let $x = r/r_e$, $E_0 = \hbar^2/mr_e^2$, $A = 2D_e e^\alpha/E_0$, $\varepsilon = 2E/E_0$, then equation (3.1) becomes

$$\left\{ \frac{d^2}{dx^2} - A\left[ e^\alpha e^{-2\alpha x} - 2\beta e^{-\alpha x} \right] + \varepsilon \right\} \chi = 0 \qquad (3.2)$$

Now, for a real parameter $\xi$, let $y = \xi e^{-\alpha x} \in [0,1]$, then (3.2) becomes

$$\left\{ y^2 \frac{d^2}{dy^2} + y\frac{d}{dy} - \frac{A}{\alpha^2}\left[ \frac{e^\alpha}{\xi^2}y^2 - 2\frac{\beta}{\xi}y \right] + \frac{\varepsilon}{\alpha^2} \right\} \chi = 0 \qquad (3.3)$$

Consider the ansatz [23] $\chi(y) = y^\mu e^{-y/2} L_n^\nu(y)$, where $L_n^\nu(y)$ is the associated Laguerre polynomial, $\mu \geq 0$, and $\nu > -1$. It is a square integrable function that is compatible with the domain of $H$, satisfies the boundary conditions and could support bound states. Inserting in (3.3) gives

$$\left\{ y\left(\frac{d}{dy} + \frac{\mu}{y} - \frac{1}{2}\right)^2 + \left(\frac{d}{dy} + \frac{\mu}{y} - \frac{1}{2}\right) - \frac{A}{\alpha^2}\left[ \frac{e^\alpha}{\xi^2}y - 2\frac{\beta}{\xi} \right] + \frac{\varepsilon}{\alpha^2 y} \right\} L_n^\nu = 0 \qquad (3.4)$$

Using the differential equation of the Laguerre polynomials, this reduces to

$$\left\{(2\mu-\nu)\frac{d}{dy}+\left(\frac{1}{4}-\frac{Ae^{\alpha}}{\alpha^{2}\xi^{2}}\right)y+\frac{1}{y}\left(\mu^{2}+\frac{\varepsilon}{\alpha^{2}}\right)+\left(2\frac{A\beta}{\alpha^{2}\xi}-n-\mu-\frac{1}{2}\right)\right\}L_{n}^{\nu}=0 \quad (3.5)$$

Now, since all terms inside the curly bracket are linearly independent then we must impose the following conditions

$$\nu=2\mu, \quad \xi^{2}=4Ae^{\alpha}/\alpha^{2}, \quad \mu=2(A\beta/\alpha^{2}\xi)-n-1/2, \text{ and } \varepsilon=-\alpha^{2}\mu^{2} \quad (3.6)$$

Giving $\mu=\frac{1}{2}\xi\beta e^{-\alpha}-n-\frac{1}{2}$ and since $\mu\geq 0$, we must have $n\leq\frac{1}{2}\xi\beta e^{-\alpha}-\frac{1}{2}$. Thus, finally the energy spectrum is found to be

$$E_{n}=-\frac{1}{2}\alpha^{2}E_{0}\left(n+\frac{1}{2}-\frac{1}{2}\xi\beta e^{-\alpha}\right)^{2} \quad (3.7)$$

which can be written more explicitly as follows

$$E_{n}=-\frac{1}{2m}(\alpha\hbar/r_{e})^{2}\left(n+\frac{1}{2}-\frac{1}{2}\gamma\beta\right)^{2}, \qquad n=0,1,..,n_{\max} \quad (3.8)$$

where $\xi=\gamma e^{\alpha}$, $\gamma=\frac{2r_{e}}{\alpha\hbar}\sqrt{2mD_{e}}$, and $n_{\max}$ is the number of bound states. It is the maximum integer that satisfies the condition $n_{\max}\leq\frac{1}{2}(\gamma\beta-1)$.

For our numerical implementation, we use the parameters that are shown in Table 1 which were obtained from [17] and [18]. Using the strategy outlined at the end of section II, we have used both the Laguerre and oscillator bases. The choice is made in favor of the basis that gives better convergence and stability for the particular molecular model. As an example, we show in Fig. 1 the behavior of a given eigenvalue (in this case $n = 10$) as a function of the basis scale parameter $\lambda$ for the $H_2$ molecule. In the Laguerre basis, a plateau is evident for values of $\lambda$ in the range 20 to 70 whereas this plateau narrows down to between 2 and 5 for the oscillator basis. Similarly, for LiH molecule we have seen a plateau for $\lambda$ in the range 40 to 100 in the Laguerre basis while in the oscillator basis it narrows down to the range 3 to 9. These observations suggest that the Laguerre basis is more suitable for both $H_2$ an LiH molecules. Nonetheless, one could still use the oscillator basis but in a narrower range of values of the parameter $\lambda$. The diagonalization of the full Hamiltonian gives the numerical bound states shown in Table 2a with the order of the eigenvalues represented by $n$ (the vibrational quantum number). These numerical computations were performed using the model parameters shown in Table 1 and in the Laguerre basis with $N = 100$ and $\lambda = 40$ for $H_2$ molecule and $N = 100$ and $\lambda = 60$ for LiH molecule. We see that these eigenvalues are in excellent agreement with the exact results computed from (3.8) up to 15 digits for low-level excitations. In the Table we had to go to higher order bound states to reach a level at which our numerical results starts deviating measurably from the exact ones. It is worth mentioning that significant increase in the accuracy can still be achieved by moderately enlarging the basis size $N$ even, say, up to $N = 150$. It is also worth noting that the final choice of the parameter $\lambda$ for a given molecule and basis is made only after studying the effect of its variation on the whole energy spectrum. This is due to the fact that the plateau for $\lambda$ narrows down as the bound state index $n$ increases while going over the whole spectrum. This behavior is demonstrated in Fig. 2 for the LiH molecule in the Laguerre basis and for $n = 0, 5, 10, 15, 20,$ and 25.

Doing the same for the other two molecules, which are included in our current study, we have the following observation. For the HCl molecule, no well-defined plateau exists for $\lambda$ in the Laguerre basis with $N = 100$ while a plateau between 3 and 8 shows up in the oscillatory basis. The same conclusion is reached for the CO molecule where a wider plateau of $\lambda$ shows up in the oscillator basis between 5 and 16 with $N = 200$. Thus, it is clear that the oscillator basis is more suitable for HCl and CO molecules whereas the Laguerre basis is more suitable for $H_2$ and HCl. Table 2b shows the numerically generated bound states for HCl and CO using the oscillator basis as compared to the exact values generated by the analytic formula (3.8). The numerical parameters were $N = 100$, $\lambda = 5$ for HCl and $N = 200$, $\lambda = 12$ for CO along with the corresponding model parameters given in Table 1. The agreement between the numerical and exact results is demonstrated up to 14 digits at low vibrations.

## IV. ROTATING MORSE POTENTIAL

In this section, we will consider the Morse potential with non-vanishing angular momentum and use our numerical procedure presented in section II to evaluate the corresponding bound states energy spectrum. The stability and the accuracy of our results depend mainly on two parameters, $\lambda$ and $N$. $N$ represents the number of basis elements that we used (dimension of our tridiagonal matrix) and $\lambda$ is the length scale parameter of the basis. The Morse potential with $\ell \neq 0$ is not exactly solvable and hence our numerical results cannot be checked against exact ones. However many numerical and perturbative results have been published in recent years [8-18]. The most widely used approximation was devised by Pekeris [4] which is based on the expansion of the centrifugal $\ell(\ell+1)/r^2$ barrier in a series of exponential terms around the equilibrium inter-nuclear position $r_e$ of the Morse potential by keeping terms up to second order in $r/r_e$ (i.e., at low excitation energy, where $r \approx r_e$). Other approximations have also been devised but they require the numerical solution of transcendental equations [7,10,13].

In Table 3 we show the bound states energy of the $H_2$ molecule for different values of the angular momentum generated by using the Laguerre basis with $\lambda = 40$, $N = 100$ and the model parameters in Table 1. We also list analogous results obtained by other approximations such as the Hypervirial perturbation method (HV) [11], the Super-Symmetric method (SUSY) [9], the Asymptotic Iteration Method (AIM) [16], the results of Duff (Duff) [6], the Modified Shifted $1/N$ expansion (MSE) [13], the Two-Point Quasi-Rational Approximation technique (TQA) [17], and the variational method using Pekeris approximation [8]. In Table 4 we give the bound states energies generated by our method for different values of the angular momentum for LiH in the Laguerre basis using the parameters $N = 100$, $\lambda = 60$ and the corresponding model parameters in Table 1. Table 5 shows the numerically generated bound states energies for HCl using the oscillator basis with $N = 100$, $\lambda = 6$ and the model parameters in Table 1. Table 6 gives the bound states energies for the CO molecule using the oscillator basis with $N = 200$, $\lambda = 12$ and using the model parameters in Table 1. The agreement between our numerical results and those generated by other methods up to four significant digits is reassuring. Again, higher accuracy could easily be achieved by increasing $N$.

## V. CONCLUSION

This work is the first in a series of articles in which we study the rotating Morse potential model for diatomic molecules in the tridiagonal J-matrix representation. Here, we provided an alternative method for obtaining the bound states energies for four different types of diatomic molecules: $H_2$, LiH, HCl and CO. Our numerical results have been compared favorably with those obtained using other approximation schemes such as the variational method using Pekeris approximation, supersymmetry, hypervirial perturbation, the shifted $1/N$ expansion, the modified shifted $1/N$ expansion, the asymptotic iteration method, and the two-point quasi-rational approximation technique. Furthermore, since realistic diatomic potentials are more accurately modeled by the perturbed or deformed Morse potentials, we are confident that our approach will produce much more accurate information about the structure and dynamics of such molecules if we use these deformed potential models. This will be the subject of investigation in one of our future articles in the series. We believe that our approach has the advantage of combining analytic and numerical powers making it very stable, rapidly converging, and highly accurate. Of course, the real advantage of our method will be exhibited when we address the issue of energy resonances while taking into account the full contribution of the reference Hamiltonian $H_0$ (even if it were to include long-range solvable potentials such as the Coulomb potential). As such, our method is systematic, highly accurate, and could easily be extended to other *short-range* potentials of which the Morse is only an example.

## ACKNOWLEDGMENTS

The authors are grateful for the support provided by the Physics department at King Fahd University of Petroleum & Minerals.

# Appendix A

In this Appendix, we complement the theoretical formulation presented in section II with the oscillator basis defined by

$$\phi_n(x) = A_n x^\alpha e^{-x/2} L_n^\nu(x); \qquad n = 0,1,2,.. \tag{A.1}$$

where $x = (\lambda r)^2$, $\alpha > 0$, $\nu > -1$, and $A_n$ is the normalization constant to be chosen below. The matrix representation of the reference Hamiltonian (2.1) $H_0$ ($\equiv H - V$) in this basis is written as

$$(H_0)_{mn} = -\frac{\lambda^2}{2} \langle \phi_m(x) | \left[ x \frac{d^2}{dx^2} + \frac{1}{2} \frac{d}{dx} - \frac{\ell(\ell+1)}{x} \right] | \phi_n(x) \rangle. \tag{A.2}$$

In the manipulation of (A.2) we use the differential equation, differential formulas to bring the matrix element in the following form

$$\frac{1}{2\lambda^2} \langle \phi_m | (H_0 - E) | \phi_n \rangle = -\left[ n(2\alpha - \nu - \tfrac{1}{2}) + (\alpha - \tfrac{1}{4})^2 - (\ell + \tfrac{1}{2})^2/4 \right] \langle \phi_m | \frac{1}{x} | \phi_n \rangle +$$
$$\left[ n + \alpha + \frac{1}{4} - \frac{E}{2\lambda^2} \right] \langle \phi_m | \phi_n \rangle - \frac{1}{4} \langle \phi_m | x | \phi_n \rangle + (n + \nu)(2\alpha - \nu - \tfrac{1}{2}) \frac{A_n}{A_{n-1}} \langle \phi_m | \frac{1}{x} | \phi_{n-1} \rangle. \tag{A.3}$$

To make this matrix representation tridiagonal we must choose the parameters such that all individual matrix terms are either tridiagonal or cancel each other for all $n$. Thus, we require $2\alpha = \nu + \tfrac{1}{2}$ and $\nu = \ell + \tfrac{1}{2}$ giving

$$\langle\phi_m|x|\phi_n\rangle = (2n+\nu+1)\delta_{n,m} - \sqrt{n(n+\nu)}\delta_{n,m+1} - \sqrt{(n+1)(n+\nu+1)}\delta_{n,m-1}, \quad \text{(A.4)}$$
$$\langle\phi_m|\phi_n\rangle = \delta_{mn}$$

where the choice of normalization, $A_n = \sqrt{\frac{2\lambda\,\Gamma(n+1)}{\Gamma(n+\nu+1)}}$, was made to obtain an orthonormal basis. Using the differential equation, differential formulas, and the three-term recursion relation of the Laguerre polynomials [22] we obtain the following elements of the matrix representation of the reference Hamiltonian

$$(H_0)_{nm} = \tfrac{\lambda^2}{2}(2n+\nu+1)\delta_{n,m} + \tfrac{\lambda^2}{2}\sqrt{n(n+\nu)}\delta_{n,m+1} + \tfrac{\lambda^2}{2}\sqrt{(n+1)(n+\nu+1)}\delta_{n,m-1} \quad \text{(A.5)}$$

Now, the only remaining quantity that is needed to perform the calculation is the matrix elements of the potential $V(r)$. This is obtained by evaluating the integral

$$V_{nm} = \int_0^\infty \phi_n(\lambda r)V(r)\phi_m(\lambda r)dr$$
$$= A_n A_m \int_0^\infty x^\nu e^{-x} L_n^\nu(x) L_m^\nu(x)\left[V(\sqrt{x}/\lambda)\right]dx \quad \text{(A.6)}$$

Using Gauss quadrature integral approximation [21], we obtain

$$V_{nm} \cong \sum_{k=0}^{N-1}\Lambda_{nk}\Lambda_{mk}\left[V(\sqrt{\varepsilon_k}/\lambda)\right] \quad \text{(A.7)}$$

where $N$ is a large enough integer. $\varepsilon_k$ and $\{\Lambda_{nk}\}_{n=0}^{N-1}$ are the respective $N$ eigenvalues and eigenvectors of the $N\times N$ tridiagonal symmetric matrix, whose elements are

$$J_{n,n} = 2n+\nu+1, \quad J_{n,n+1} = -\sqrt{(n+1)(n+\nu+1)}. \quad \text{(A.7)}$$

The difference between this quadrature matrix, which is associated with the oscillator basis, and that in Eq. (2.8), which is associated with the Laguerre basis, is that here we use $\nu = \ell + \tfrac{1}{2}$ not $\nu = 2\ell+1$. Moreover, one should make note of the marked difference in the term within square brackets in Eq. (A.7) as opposed to that in Eq. (2.7) for the Laguerre basis.

**REFERENCES:**

[1] P. M. Morse, Phys. Rev. **34** (1929) 57
[2] S. H. Dong, R. Lemus, and A. Frank, Int. J. Quant. Phys. **86** (2002) 433, and references therein
[3] S. Flügge, *Practical Quantum Mechanics*, vol. I (Springer, Berlin, 1994)
[4] C. L. Pekeris, Phys. Rev. **45** (1934) 98
[5] R. Herman and R. J. Rubin, Astrophys. J. **121** (1955) 533
[6] M. Duff and H. Rabitz, Chem. Phys. Lett. **53** (1978) 152
[7] J. R. Elsum and G. Gordon, J. Chem. Phys. **76** (1982) 5452
[8] E. D. Filho and R. M. Ricotta, Phys. Lett. A **269** (2000) 269
[9] F. Cooper, A. Khare, and U. Sukhatme, Phys. Rep. **251** (1995) 267
[10] D. A. Morales, Chem. Phys. Lett. **394** (2004) 68
[11] J. P. Killingbeck, A. Grosjean, and G. Jolicard, J. Chem. Phys. **116** (2002) 447
[12] T. Imbo and U. Sukhatme, Phys. Rev. Lett. **54** (1985) 2184
[13] M. Bag, M. M. Panja, R. Dutt, and Y. P. Varshni, Phys. Rev. A **46** (1992) 6059
[14] A. F. Nikiforov and V. B. Uvarov, *Special Functions of Mathematical Physics* (Birkhausr, Basel, 1988); C. Berkdemir and J. Han, Chem. Phys. Lett. **409** (2005) 203
[15] C. Berkdemir, Nucl. Phys. A **770** (2006) 32
[16] O. Bayrak and I. Boztosum, J. Phys. A **39** (2006) 6955
[17] E. Castro, J. L. Paz and P. Martin, J. Mol. Str.: THEOCHEM **769** ( 2006) 15
[18] W. C. Qiang and S. H. Dong, Phys. Lett. A **363** (2006) 169
[19] C. C. Marston and G. G. Balint-Kurti, J. Chem. Phys. **91** (1989) 3571
[20] E. J. Heller and H. A. Yamani, Phys. Rev. A **9** (1974) 1201; H. A. Yamani and L. Fishman, J. Math. Phys. **16** (1975) 410; A. D. Alhaidari, E. J. Heller, H. A. Yamani, and M. S. Abdelmonem (eds.), *The J-matrix Method: Recent developments and selected applications* (Springer, Heidelberg, 2007)
[21] See, for example, Appendix A in: A. D. Alhaidari, H. A. Yamani, and M. S. Abdelmonem, Phys. Rev. A **63** (2001) 062708
[22] G. Szegö, *Orthogonal polynomials*, 4th ed. (Am. Math. Soc., Providence, RI, 1997); T. S. Chihara, *An Introduction to Orthogonal Polynomials* (Gordon and Breach, New York, 1978); N. I. Akhiezer, *The Classical Moment Problem* (Oliver and Boyd, Einburgh, 1965); R. W. Haymaker and L. Schlessinger, *The Padé Approximation in Theoretical Physics*, edited by G. A. Baker and J. L. Gammel (Academic, New York, 1970); D. G. Pettifor and D. L. Weaire (eds), *The Recursion Method and its Applications* (Springer, Berlin, 1985)
[23] A. D. Alhaidari, Ann. Phys. **317** (2005) 152: Sec. 5, Eq. (5.7)

**TABLE CAPTIONS:**

**Table 1**: Model parameters for the diatomic molecules in our study as obtained from the cited sources. $E_0$ is a derived quantity, which is calculated as $E_0 = \hbar^2/m\, r_e^2$.

**Table 2**: Numerical and exact bound states energies ($-E$) for the S-wave Morse potential ($\ell = 0$). The Laguerre (Oscillator) basis was used for $H_2$ and LiH (HCl and CO) molecules. The integer $n$ represents the energy level in the spectrum. Numerically $n_{max}$ is the maximum index of the energy level beyond which the eigenvalues change sign.

**Table 3**: Bound states energy eigenvalues ($-E$) for the $H_2$ molecule (in eV) for different values of the rotational $\ell$ and vibrational $n$ quantum numbers in the Laguerre basis with $N = 100$ and $\lambda = 40$.

**Table 4**: Bound states energy eigenvalues ($-E$) for the LiH molecule (in eV) for different values of the rotational $\ell$ and vibrational $n$ quantum numbers in the Laguerre basis with $N = 100$ and $\lambda = 60$.

**Table 5**: Bound states energy eigenvalues ($-E$) for the HCl molecule (in eV) for different values of the rotational $\ell$ and vibrational $n$ quantum numbers in the Oscillator basis with $N = 100$ and $\lambda = 6$.

**Table 6**: Bound states energy eigenvalues ($-E$) for the CO molecule (in eV) for different values of the rotational $\ell$ and vibrational $n$ quantum numbers in the Oscillator basis with $N = 100$ and $\lambda = 12$.

**FIGURE CAPTIONS:**

**Fig. 1:** Variations in the computed value of one of the bound states energy levels of $H_2$ molecule with the basis scale parameter $\lambda$. A plateau of computational stability is evident for the range $\lambda = 20$ to 70 ($\lambda = 2$ to 5) in the Laguerre (Oscillator) basis. We took $\ell = 0$ and basis size $N = 50$.

**Fig. 2:** Same as Fig. 1, but for a few energy levels from across the whole spectrum. We took $\ell = 0$ and used the Laguerre basis with basis size $N = 100$. For better display, we shifted each energy level by a constant to bring it to line with the rest.

**Fig. 3:** A plot of the square root of the bound states energy vs. the vibrational quantum number for the LiH molecule and for different angular momenta. The Laguerre basis was used with $\lambda = 60$ and $N = 100$. The units on the vertical axis is $eV^{1/2}$.

**Fig. 4:** The bound states energy vs. the angular momentum quantum number for the LiH molecule and for different vibrational numbers. The Laguerre basis was used with $\lambda = 70$ and $N = 50$.

**Table 1**

| Molecule | $D_e$ (eV) | $r_e$ (Å) | $m$ (amu) | $\alpha$ | $E_0$ (eV) |
|---|---|---|---|---|---|
| $H_2$ [17] | 4.7446 | 0.7416 | 0.50391 | 1.440558 | $1.508343932\times10^{-2}$ |
| LiH [18] | 2.515287 | 1.5956 | 0.8801221 | 1.7998368 | $1.865528199\times10^{-3}$ |
| HCl [18] | 4.61907 | 1.2746 | 0.9801045 | 2.38057 | $2.625261613\times10^{-3}$ |
| CO [18] | 11.2256 | 1.1283 | 6.8606719 | 2.59441 | $4.786047154\times10^{-4}$ |

**Table 2a**

| | $H_2$ | | LiH | |
|---|---|---|---|---|
| $n$ | This work | Exact | This work | Exact |
| 0 | 4.476013136943936 | 4.476013136943926 | 2.428863215520034 | 2.428863215520037 |
| 1 | 3.962315358958284 | 3.962315358958260 | 2.260548058095126 | 2.260548058095136 |
| 2 | 3.479918845141218 | 3.479918845141241 | 2.098276116050267 | 2.098276116050265 |
| 3 | 3.028823595492864 | 3.028823595492867 | 1.942047389385409 | 1.942047389385426 |
| 4 | 2.609029610013135 | 2.609029610013140 | 1.791861878100625 | 1.791861878100618 |
| 5 | 2.220536888702045 | 2.220536888702059 | 1.647719582195847 | 1.647719582195841 |
| 6 | 1.863345431559636 | 1.863345431559624 | 1.509620501671092 | 1.509620501671094 |
| 7 | 1.537455238585834 | 1.537455238585836 | 1.377564636526370 | 1.377564636526379 |
| 8 | 1.242866309780691 | 1.242866309780693 | 1.251551986761669 | 1.251551986761694 |
| 9 | 0.979578645144206 | 0.979578645144197 | 1.131582552377003 | 1.131582552377041 |
| 10 | 0.747592244676326 | 0.747592244676347 | 1.017656333372503 | 1.017656333372418 |
| 11 | 0.546907108377146 | 0.546907108377143 | 0.909773329748332 | 0.909773329747826 |
| 12 | 0.377523236246581 | 0.377523236246586 | 0.807933541501811 | 0.807933541503265 |
| 13 | 0.239440628284702 | 0.239440628284674 | 0.712136968622215 | 0.712136968638735 |
| 14 | 0.132659284491403 | 0.132659284491409 | 0.622383611097279 | 0.622383611154236 |
| 20 | | | 0.210770988653686 | 0.210770989227890 |
| 25 | | | 0.033948880578186 | 0.033948893906786 |
| $n_{max}$ | 17 | 17 | 29 | 29 |
| $N, \lambda$ | 100, 40 | | 100, 60 | |
| Basis | Laguerre | | Laguerre | |

**Table 2b**

| | HCl | | CO | |
|---|---|---|---|---|
| $n$ | This work | Exact | This work | Exact |
| 0 | 4.435563943352675 | 4.435563943352696 | 11.09153532340909 | 11.09153532340921 |
| 1 | 4.079710071754065 | 4.079710071754069 | 10.82582207328953 | 10.82582207328966 |
| 2 | 3.738733855750068 | 3.738733855750079 | 10.56333029391945 | 10.56333029391951 |
| 3 | 3.412635295340741 | 3.412635295340728 | 10.30405998529871 | 10.30405998529875 |
| 4 | 3.101414390526035 | 3.101414390526015 | 10.04801114742742 | 10.04801114742739 |
| 5 | 2.805071141305942 | 2.805071141305941 | 9.795183780305372 | 9.795183780305433 |
| 6 | 2.523605547680502 | 2.523605547680504 | 9.545577883932884 | 9.545577883932868 |
| 7 | 2.257017609649704 | 2.257017609649706 | 9.299193458309665 | 9.299193458309699 |
| 8 | 2.005307327213539 | 2.005307327213546 | 9.056030503435888 | 9.056030503435929 |
| 9 | 1.768474700372031 | 1.768474700372025 | 8.816089019311511 | 8.816089019311555 |
| 10 | 1.546519729125149 | 1.546519729125141 | 8.579369005936609 | 8.579369005936579 |
| 11 | 1.339442413472914 | 1.339442413472896 | 8.345870463311027 | 8.345870463310998 |
| 12 | 1.147242753415306 | 1.147242753415289 | 8.115593391434810 | 8.115593391434814 |
| 13 | 0.969920748952278 | 0.969920748952320 | 7.888537790307976 | 7.888537790308028 |
| 14 | 0.807476400083812 | 0.807476400083989 | 7.664703659930634 | 7.664703659930638 |
| 18 | 0.306475560556244 | 0.306475560557049 | 6.801581845915044 | 6.801581845915046 |
| 21 | 0.086940302112394 | 0.086940314655545 | 6.188065928271975 | 6.188065928272019 |
| 40 | | | 2.975752503156747 | 2.975752503156777 |
| 60 | | | 0.850620702617658 | 0.850743542668641 |
| $n_{max}$ | 24 | 25 | 70 | 83 |
| $N, \lambda$ | 100, 5 | | 200, 12 | |
| Basis | Oscillator | | Oscillator | |

**Table 3**

| $n$ | $\ell$ | This work | HV [11] | SUSY [10] | AIM [16] | Duff [6] | MSE [13] | Variational [8] | TQA [17] |
|---|---|---|---|---|---|---|---|---|---|
| 0 | 0  | 4.4760131 | 4.47601 | 4.47601 | 4.47601 | 4.4762 | 4.4760 | 4.4758 | 4.4760084 |
|   | 5  | 4.2590180 | 4.25901 | 4.25880 | 4.25880 | 4.2592 | 4.2590 | 4.2563 | 4.2590038 |
|   | 10 | 3.7247471 | 3.72473 | 3.72193 | 3.72193 | 3.7251 | 3.7247 | 3.7187 | 3.7247181 |
|   | 15 | 2.9663814 | 2.96635 | 2.95158 |         | 2.9669 |        | 2.9578 | 2.9663319 |
|   | 20 | 2.0840636 | 2.08401 | 2.02864 |         |        |        | 2.0735 | 2.0839937 |
| 5 | 0  | 2.2205369 | 2.22051 | 2.22051 | 2.22052 | 2.218  | 2.2205 |        |           |
|   | 5  | 2.0528808 | 2.05285 | 2.04353 | 2.04355 | 2.050  | 2.0430 |        |           |
|   | 10 | 1.6526902 | 1.65265 | 1.60389 | 1.60391 | 1.650  | 1.6535 |        |           |
|   | 15 | 1.1117173 | 1.11167 | 0.96626 |         | 1.110  |        |        |           |
|   | 20 | 0.5237657 | 0.52372 | 0.18940 |         |        |        |        |           |
| 6 | 0  | 1.8633454 | 1.86332 | 1.86332 |         |        |        |        |           |
|   | 5  | 1.7069051 | 1.70687 | 1.69439 |         |        |        |        |           |
|   | 10 | 1.3363630 | 1.33633 | 1.27418 |         |        |        |        |           |
|   | 15 | 0.8422858 | 0.84224 | 0.66310 |         |        |        |        |           |
|   | 20 | 0.3179828 | 0.31794 | −0.08454|         |        |        |        |           |
| 7 | 0  | 1.5374552 | 1.53743 | 1.53743 | 1.53744 |        | 1.5374 |        |           |
|   | 5  | 1.3926614 | 1.39263 | 1.37654 | 1.37656 |        | 1.3932 |        |           |
|   | 10 | 1.0526836 | 1.05265 | 0.97578 | 0.97581 |        | 1.0552 |        |           |
|   | 15 | 0.6067737 | 0.60673 | 0.39125 |         |        |        |        |           |
|   | 20 | 0.1487217 | 0.14869 | −0.32718|         |        |        |        |           |

**Table 4**

| $n$ | $\ell$ | This work | EQR [18] | SUSY [10] | FGH [19] | AIM [16] | MSE [13] | NU [15] | Variational [8] |
|---|---|---|---|---|---|---|---|---|---|
| 0 | 0  | 2.4288627 | 2.42886 | 2.42886 | 2.42886 | 2.4289 | 2.4280 | 2.4287 | 2.4291 |
|   | 5  | 2.4013352 | 2.40133 | 2.40133 | 2.40133 | 2.4013 | 2.4000 | 2.4012 | 2.4014 |
|   | 10 | 2.3288530 | 2.32883 | 2.32883 | 2.32885 | 2.3288 | 2.3261 | 2.3287 | 2.3287 |
|   | 15 | 2.2138464 | 2.21377 | 2.21377 | 2.21385 |        |        |        |        |
|   | 20 | 2.0600073 | 2.05977 | 2.05977 | 2.06001 |        |        |        |        |
| 5 | 0  | 1.6477149 | 1.64772 | 1.64772 | 1.64772 | 1.6477 | 1.6402 | 1.6476 |        |
|   | 5  | 1.6239497 | 1.62377 | 1.62377 | 1.62395 | 1.6238 | 1.6160 | 1.6236 |        |
|   | 10 | 1.5615114 | 1.56074 | 1.56074 | 1.56152 | 1.5607 | 1.5525 | 1.5606 |        |
|   | 15 | 1.4628514 | 1.46079 | 1.46078 | 1.46286 |        |        |        |        |
|   | 20 | 1.3316742 | 1.32718 | 1.32718 | 1.33168 |        |        |        |        |
| 7 | 0  | 1.3775588 | 1.37757 | 1.37756 | 1.37756 | 1.3776 | 1.3862 | 1.3774 |        |
|   | 5  | 1.3553770 | 1.35505 | 1.35505 | 1.35538 | 1.3550 | 1.3456 | 1.3549 |        |
|   | 10 | 1.2971612 | 1.29581 | 1.29580 | 1.29715 | 1.2958 | 1.2865 | 1.2956 |        |
|   | 15 | 1.2053616 | 1.20190 | 1.20189 | 1.20533 |        |        |        |        |
|   | 20 | 1.0836775 | 1.07645 | 1.07644 | 1.08350 |        |        |        |        |

**Table 5**

| $n$ | $\ell$ | This work | EQR [18] | SUSY [10] | FGH [19] | AIM [16] | MSE [13] | Variational [8] |
|---|---|---|---|---|---|---|---|---|
| 0 | 0 | 4.4355522 | 4.43556 | 4.43556 | 4.43556 | 4.4356 | 4.4355 | 4.4360 |
|   | 5 | 4.3968066 | 4.39681 | 4.39681 | 4.39682 | 4.3968 | 4.3968 | 4.3971 |
|   | 10 | 4.2940628 | 4.28407 | 4.28408 | 4.28409 | 4.2841 | 4.2940 | 4.2840 |
|   | 15 | 4.1288846 | 4.12889 | 4.12889 | 4.12893 |   |   |   |
|   | 20 | 3.9037744 | 3.90374 | 3.90375 | 3.90385 |   |   |   |
| 5 | 0 | 2.8049687 | 2.80507 | 2.80508 | 2.80508 | 2.8051 | 2.8046 |   |
|   | 5 | 2.7721880 | 2.77210 | 2.77211 | 2.77230 | 2.7721 | 2.7718 |   |
|   | 10 | 2.6853673 | 2.68472 | 2.68473 | 2.68549 | 2.6847 | 2.6850 |   |
|   | 15 | 2.5461184 | 2.54442 | 2.54443 | 2.54626 |   |   |   |
|   | 20 | 2.3570303 | 2.35354 | 2.35355 | 2.35719 |   |   |   |
| 7 | 0 | 2.2568924 | 2.25702 | 2.25703 | 2.25703 | 2.2570 | 2.2565 |   |
|   | 5 | 2.2265969 | 2.22636 | 2.22636 | 2.22673 | 2.2263 | 2.2262 |   |
|   | 10 | 2.1464148 | 2.14512 | 2.14513 | 2.14656 | 2.1451 | 2.1461 |   |
|   | 15 | 2.0179894 | 2.01477 | 2.01478 | 2.01815 |   |   |   |
|   | 20 | 1.8439683 | 1.83760 | 1.83761 | 1.84415 |   |   |   |

**Table 6**

| $n$ | $\ell$ | This work | EQR [18] | SUSY [10] | FGH [19] | AIM [16] | MSE [13] | NU [15] | Variational [8] |
|---|---|---|---|---|---|---|---|---|---|
| 0 | 0 | 11.0915353 | 11.0915 | 11.0915 | 11.0915 | 11.0915 | 11.092 | 11.091 | 11.093 |
|   | 5 | 11.0843875 | 11.0844 | 11.0844 | 11.0844 | 11.0845 | 11.084 | 11.084 | 11.084 |
|   | 10 | 11.0653334 | 11.0653 | 11.0653 | 11.0653 | 11.0653 | 11.065 | 11.065 | 11.0653 |
|   | 15 | 11.0343911 | 11.0344 | 11.0344 | 11.0344 |   |   |   |   |
|   | 20 | 10.9915902 | 10.9916 | 10.9916 | 10.9916 |   |   |   |   |
| 5 | 0 | 9.7951838 | 9.79519 | 9.79519 | 9.79519 | 9.7952 | 9.795 | 9.795 |   |
|   | 5 | 9.7883443 | 9.78835 | 9.78834 | 9.78835 | 9.7883 | 9.788 | 9.788 |   |
|   | 10 | 9.7701124 | 9.77011 | 9.77010 | 9.77011 | 9.7701 | 9.770 | 9.769 |   |
|   | 15 | 9.7405064 | 9.74049 | 9.74049 | 9.74051 |   |   |   |   |
|   | 20 | 9.6995563 | 9.69952 | 9.69952 | 9.69956 |   |   |   |   |
| 7 | 0 | 9.2991935 | 9.29920 | 9.29920 | 9.29920 | 9.2992 | 9.299 | 9.299 |   |
|   | 5 | 9.2924786 | 9.29248 | 9.29248 | 9.29248 | 9.2925 | 9.292 | 9.292 |   |
|   | 10 | 9.2745791 | 9.27457 | 9.27458 | 9.27458 | 9.2745 | 9.274 | 9.274 |   |
|   | 15 | 9.2455135 | 9.24548 | 9.24548 | 9.24552 |   |   |   |   |
|   | 20 | 9.2053118 | 9.20534 | 9.20534 | 9.20531 |   |   |   |   |

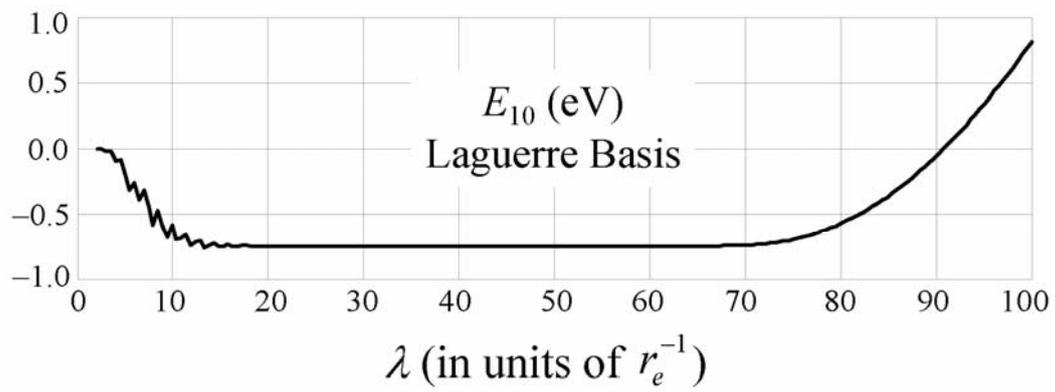

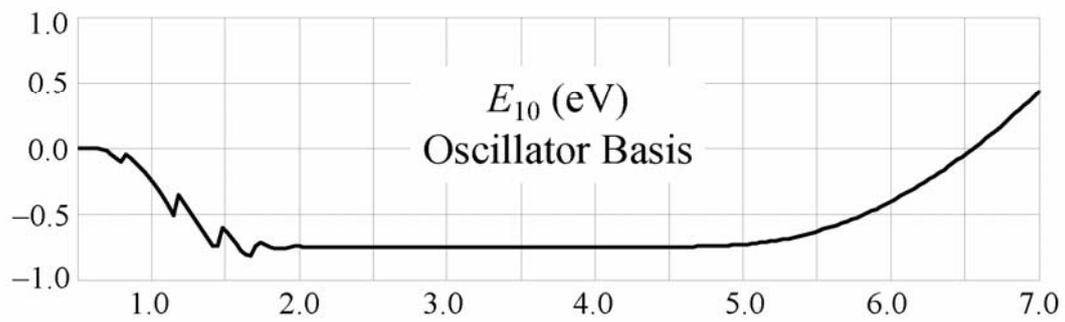

**Fig. 1**

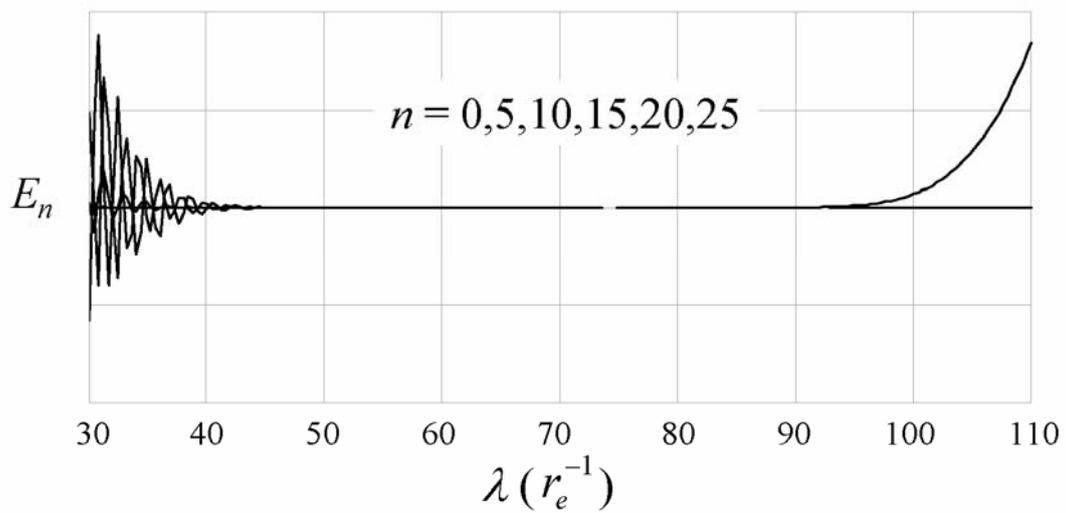

**Fig. 2**

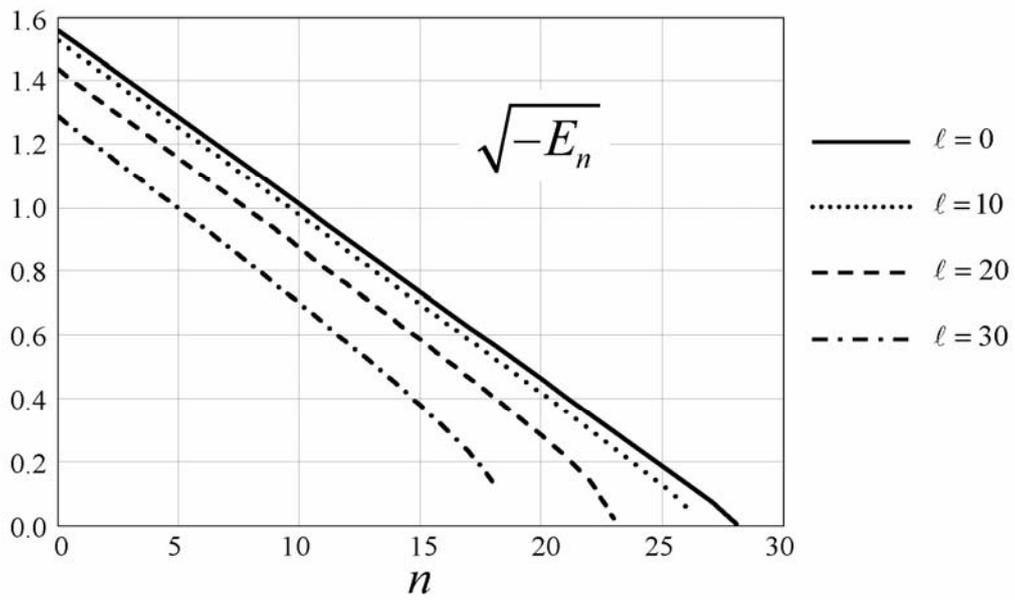

**Fig. 3**

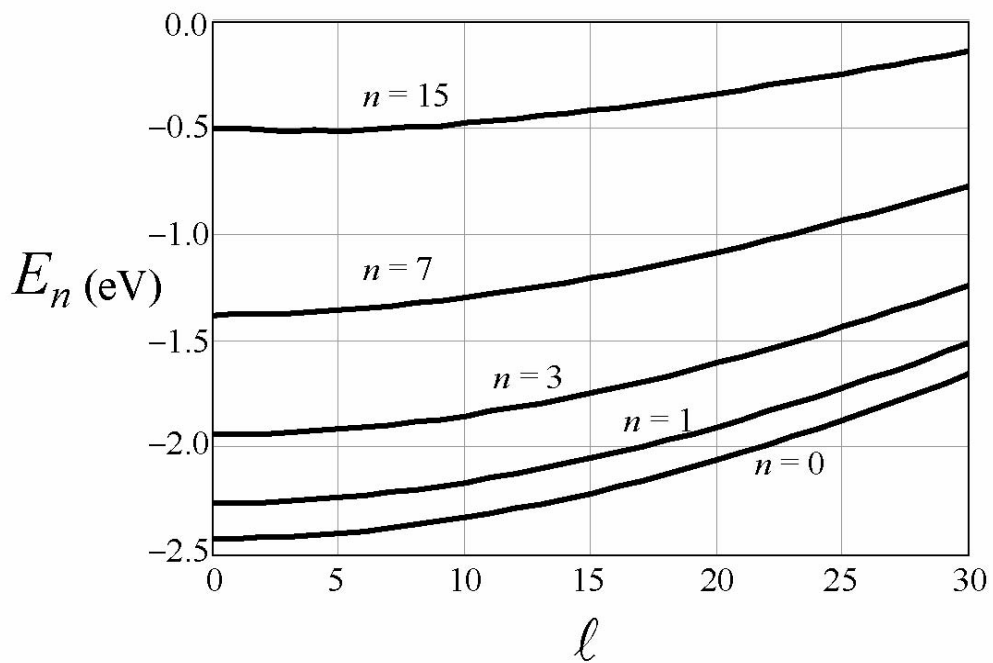

**Fig. 4**